\documentclass{ws-p8-50x6-00}

\begin{document}

\title{Neutrino Oscillation Experiments\\
             at Nuclear Reactors}

\author{Giorgio Gratta}

\address{Physics Department, Stanford University, Stanford, CA 94305, USA \\
         E-mail: gratta@hep.stanford.edu}

\maketitle

\abstracts{
In this paper I give an overview of the status of neutrino oscillation 
experiments performed using nuclear reactors as sources of neutrinos.
I review the present generation of experiments (Chooz and Palo Verde)
with baselines of about 1~km as well as the next generation 
that will search for oscillations with a baseline of about 100~km.
While the present detectors provide essential input towards the understanding
of the atmospheric neutrino anomaly, in the future, the KamLAND reactor 
experiment represents our best opportunity to study very small mass neutrino mixing
in laboratory conditions.  In addition 
KamLAND with its very large fiducial mass
and low energy threshold, will also be sensitive to a broad range of different
physics.}

\section{Introduction}

Neutrino oscillations, if discovered, would shed light on some of the 
most essential issues of modern particle physics ranging from a better 
understanding of lepton masses to the exploration of new physics beyond    
the Standard Model.    In addition finite neutrino masses would
have important consequences in astrophysics and cosmology.

Experiments performed using both particle accelerators and
nuclear reactors have been carried-on extensively in the past 20 years
finding no firm evidence for neutrino oscillations. However, in recent
years, evidence has been collected on a number of effects that could point to
oscillations: the solar neutrino puzzle, the anomaly observed in
atmospheric neutrinos\cite{ysuzuki} and the LSND effect\cite{Louis}. 
This paper will concentrate  on the first two cases that are well suited
to be studied with reactor experiments.
Both effects, if interpreted 
as signals for neutrino oscillations,
would suggest very small neutrino mass differences and, possibly,
large mixing parameters.    
We write the probability of oscillation from a flavor $\ell$ to another
one $\ell^{\prime}$ as
\begin{equation}
  P_{\ell\ell^{\prime}} = \sin^2{2\theta}
  \sin^2{{1.27 \Delta m^2 L}\over{E_{\nu}}} 
\end{equation}
\noindent where $L$ is expressed in meters, $\Delta m^2$ in eV$^2$ and 
$E_{\nu}$ in MeV.  It is clear that in order to probe sufficiently
small $\Delta m ^2$, long baselines have to be combined with low energy
neutrinos.
Unfortunately we are able to collimate neutrino beams only by using
the Lorentz boost of the parent particles from which decay the neutrinos
are produced.  
For this reason low energy neutrinos are generally
produced over large solid angles while high energy ones may come in 
relatively narrow beams.   Hence to access, for instance, the atmospheric
neutrino $\Delta m^2$ region
we have the choice of either using the beam from an accelerator
that is rather narrow (better than $\approx 1$~mrad) but has an energy of 
several GeV, or detecting few-MeV neutrinos emitted isotropically 
by a nuclear reactor.   In the first case the baseline
will have to be much larger, but since the beam is pointing, both cases
turn out to be quite feasible and their different features make them quite
complementary.
As reactors produce exclusively electron anti-neutrinos,
only $\bar\nu_{\rm e} - \bar\nu_{\rm X}$ oscillations can be observed.   
In addition since the neutrino energy is below the 
threshold for producing muons (or $\tau$s), reactor experiments have to
be of ``disappearance'' type, that is oscillation would be detected 
as a deficit of electron neutrinos.  This feature, together with the 
higher energies produced in accelerator neutrino events and their 
time-bunched structure, makes accelerators-based experiments more
immune to backgrounds and, in general, more sensitive to small
mixing parameters.   On the other hand the very low energy of reactor
neutrinos offer the best chance to push to the limit our exploration
of the small $\Delta m^2$ regime.
   
Two reactor-based experiments have been performed to study the parameter
region consistent with atmospheric neutrinos extending our reach in $\Delta m^2$
by over an order of magnitude.    While the Chooz~\cite{chooz} experiment has 
been completed (although part of the data is still being analyzed), 
Palo Verde~\cite{paloverde}  is taking data since the fall 1998 and will 
continue at least until the end of 1999.  KamLAND\cite{kamland} will 
be the first laboratory-style experiment able to 
attack one of the regions of solar neutrino oscillations.

\section{Reactors as Neutrino Sources}

Nuclear reactors produce isotropically $\bar\nu_{\rm e}$ in the 
$\beta$ decay of the
neutron-rich fission fragments.  For all practical purposes the neutrino 
flux and spectrum depend only on the composition of the core in terms
of the four isotopes $^{235}$U, $^{238}$U, $^{239}$Pu and $^{241}$Pu
being fissioned in the reactor.   Neutrinos are then produced by long
chains of daughter isotopes and hundreds of different $\beta$-decays 
have to be included to account for the observed yields.   The modeling
of such processes is quite a formidable task but there is nowadays a 
very good agreement between theoretical calculations and experimental 
data.  Two methods can be used to experimentally 
cross check theoretical models.
In one case the electron spectra for fission-produced chains can be 
experimentally measured for each of the four parent isotopes. From this data, 
available only for ($^{235}$U, $^{239}$Pu and $^{241}$Pu), anti-neutrino
spectra can be derived without loss of accuracy\cite{schre}, obtaining
a total uncertainty on the flux of about 3\%.
Alternatively anti-neutrino flux and spectra have been directly measured
\cite{oldexp} in several high statistic experiments with detectors of known 
efficiency.   These data are usually a by-product of previous reactor 
oscillation experiments where the anti-neutrino 
have been measured at different distances.  Since  
these observations have been found to be consistent with a $1/r^2$ law 
(no oscillations at short baselines) they can now be used as a determination 
of the absolute anti-neutrino spectra.  A total error of about 1.4\% has 
been achieved in these measurements.

All measurements and calculations agree with each other within
errors so that, given the history of power and fuel composition for a 
reactor, its anti-neutrino energy spectrum can be computed with an error
of about 3\%.
We note here that for this kind of experiments a ``near measurement'' is superfluous 
as, in essence, all the information needed can be readily derived from the previous 
generation of experiments, using their result that no oscillations take
place at those shorter baselines. The real challenges consist in measuring
precisely the detector efficiency and in subtracting backgrounds properly.

Since the neutrino spectrum is only measured above some energy threshold
($M_{\rm n} - M_{\rm p} + m_{\rm e}$ = 1.8 MeV), only 
fast (energetic) decays contribute to the useful flux and the ``neutrino
luminosity'' tracks very well in time the power output of the reactor.
Generally few hours after a reactor turns off the neutrino flux above
threshold has become negligible.    Similarly, the equilibrium
for neutrinos above threshold is established already several hours
after the reactor is turned on.

While early oscillation experiments used military or research reactors,
modern experiments have long baselines and so need the largest available
fluxes (powers) that are usually available at large commercial power 
generating stations.
Typical modern reactors have thermal power in excess of 3~GW ($>1$~GW 
electrical power) corresponding to $\simeq 7.7\times 10^{20}\nu/{\rm s}$.
Usually more than one such reactors are located next to each other in a 
power plant so that the neutrino flux detected is the sum of the
contributions from each core.
\begin{figure}[htbp]
\begin{center}
\fbox{\epsfig{file=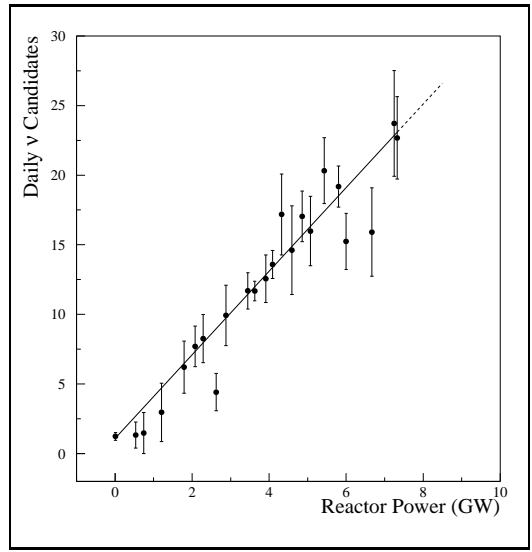,height=7cm,clip=}}
\caption{Neutrino candidates in the Chooz detector as function of the 
thermal power in GW.}
\label{fig:chooz_turn_on}
\end{center}
\end{figure}
Although periods of time with source off would be very useful to study the 
backgrounds, in the case of multiple reactors, plant optimization requires 
the refueling of one reactor at the time so that in practice backgrounds are
often studied at partial power (instead of zero power).
Typically each reactor is off (refueling) for about one month
every one or two years.   A notable exception is Chooz where the experiment was
running before the power plant was operational.    This experiment was then
able to record the slow turn on of the reactors during commissioning as shown 
in Figure~\ref{fig:chooz_turn_on}.
This is used to cross check other estimates of the backgrounds to the measurement.
\begin{figure}[htbp]
\begin{center}
\mbox{\epsfig{file=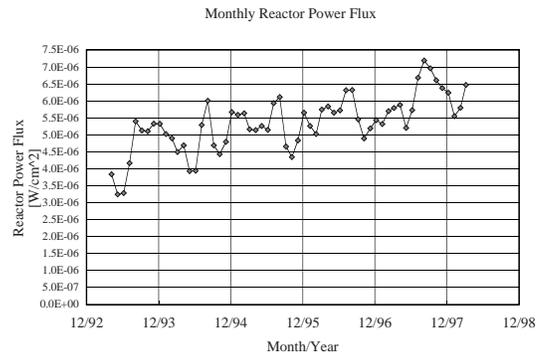,height=7cm,clip=,angle=90}}
\caption{Time-modulation of the anti-neutrino flux from reactors at
Kamioka.  The coherent decrease of flux in the spring and fall is due to
reactor refueling and maintenance performed when electricity demand is 
at minimum.}
\label{fig:flux_mod}
\end{center}
\end{figure}
KamLAND will detect neutrinos from a very large number of reactors in several 
power plants distributed in the central region of Japan.   In this case an
important check of backgrounds will result from the study of a 6-month period
modulation in the neutrino flux due to concentrated reactor refuelings and
maintenance in the fall and spring periods, when electricity demand is lowest.
Such a modulation, with a strength of about 30\% of the full flux, is
illustrated in Figure~\ref{fig:flux_mod}.

\section{Oscillation Searches in the Atmospheric Neutrino Region}

At the time of writing two experiments are exploring the region of phase-space
with  $10^{-3} < \Delta m^2 < 10^{-2}$: Chooz in France (a 2-reactor site)
and Palo Verde in the United States (a 3-reactor site).   
In order to be sensitive to oscillations in the atmospheric neutrino region
the Chooz and Palo Verde detectors are located, respectively, $\simeq$1~km and 
$\simeq$0.8~km from the reactors.
In both cases the detection is based on the inverse-$\beta$ reaction 
\begin{equation}
\bar{\nu_e} + p \rightarrow e^{+} + n.
\end{equation}
in Gadolinium-loaded liquid scintillator.
The detectors can measure the positron energy so that the 
anti-neutrino spectrum can be easily reconstructed from simple kinematics as
\begin{equation}
  E_{\bar{\nu}} = E_{{\rm e}^+} + (M_{\rm n} - M_{\rm p} + m_{\rm e}) + 
                  {\mathcal O}(E_{\bar{\nu}}/M_{\rm n}).
\end{equation}

Given a fixed baseline, different energies have different oscillation
probabilities and, for a large range of 
$\Delta m ^2$ values, the signature of oscillations is an unmistakable 
distortion of the energy spectrum.
The slight difference in baselines for the two experiments could result in
rather different oscillation signals, providing a nice cross-check against
non-oscillation effects. 
Parameter sets closer to the sensitivity boundaries will ultimately
give neutrino spectra similar to the case of no oscillations, so that
to reach the best sensitivity both experiments will have to rely  
on the absolute neutrino flux measurement.
Since at these large distances from the reactors the flux of neutrinos is rather 
low, special precautions have to be taken in order to suppress backgrounds from 
cosmic radiation and natural radioactivity.
Although both detectors are located underground, background rejection is achieved 
somewhat differently in the two cases. On one hand Chooz has been 
built in a rather deep (300 m.w.e.) already existing underground site, 
while, on the other, Palo Verde was installed in a shallow laboratory (32 m.w.e.) 
excavated on purpose. Hence this last experiment is segmented and uses 
tighter signatures to identify anti-neutrino events. 

The central detector of Palo Verde, a matrix of 66 acrylic cells each 9 m long, is
surrounded by a 1 m thick water buffer that shields $\gamma$ radiation 
and neutrons.   A large veto counter encloses the entire detector, 
rejecting cosmic-ray muons.
In this detector the signal consists of a fast triple coincidence 
followed by the neutron capture. The triple is produced
by the ionization due to the positron and, in two different cells, its 
two annihilation photons.   Timing information at the two ends of each 
cell allows to reconstruct the events longitudinally and to correct for 
light attenuation in the cells, providing a good quality positron energy 
measurement.

The Chooz detector, on the other hand, being in a lower background environment,
is a single spherical acrylic vessel filled with liquid scintillator.
It triggers on the double coincidence between the positron and the neutron 
parts of the inverse $\beta$ reactions.  
Also in this case the central detector is surrounded by a veto and some
shielding layers.

Gadolinium doping of the scintillator reduces the neutron capture time  
and hence the background.  A concentration of 0.1\% Gd by weight 
reduces the capture time from 170$\mu$s to 28$\mu$s.
Since a neutron capture in Gd is accompanied by a 8 MeV photon cascade,
another advantage of the doped scintillator is that it allows for a very high
threshold for the neutron part of the event.   This threshold, well above
the Th and U lines, results in further reduction of the background.

Although both detectors are built using low activity materials, this requirement
is more severe for Chooz in order to have a $\gamma$-ray rate
consistent with the lower cosmic-ray induced background.

Two categories of backgrounds are considered: one is given by random hits in the
detector (2 for Chooz, 4 for Palo Verde) produced by independent $\gamma$-rays
and/or neutrons, while the other is given by single or double fast neutrons produced
outside the veto by cosmic-ray muons mainly in spallation processes.   Neutrons can
deposit some energy simulating the fast part of the event  
and then thermalize and capture in Gd (like the neutrons from the anti-neutrino
capture process).   Unlike the case of independent hits, in this second background the event
has the same time-structure of real events, so that its rejections is 
a-priori more difficult.
The expected rates of neutrino events for the case of no oscillations is
about 30 day$^{-1}$ for both detectors.
Both groups use rather advanced trigger and data acquisition systems to
select and log neutrino events.

\begin{figure}[htbp]
\begin{center}
\mbox{\epsfig{file=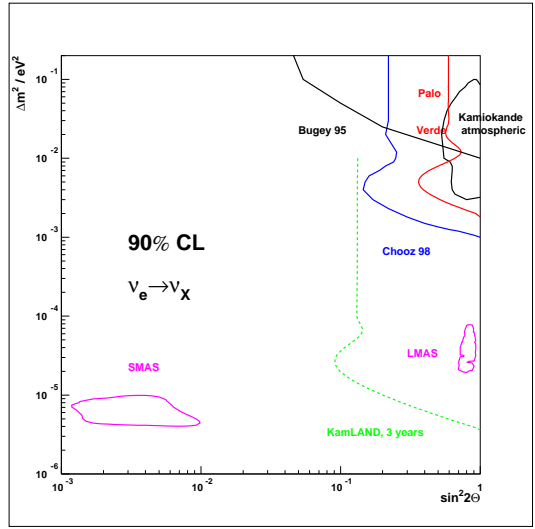,height=7cm,clip=}}
\caption{Phasespace for $\nu_{\rm e} - \nu_{\mu}$ oscillations in the 
 region relevant for modern reactor experiments.
 The 90\% CL limits from Chooz and Palo Verde are shown together with
 the KamLAND projected sensitivity assuming 3 years of data-taking and
 no evidence for oscillations.  The Palo Verde result includes only the 
 first 70 days of data in 1998. The MSW mechanism is used in plotting the
 solar neutrino regions.   The sensitivity of reactor experiments is the same
 for $\nu_{\rm e} - \nu_{\tau}$ oscillations.}
\label{fig:excl_summary}
\end{center}
\end{figure}

At the present time both experiments see fluxes that are completely compatible with the 
expectations for no oscillations.
From these observations the reactor experiments can exclude that oscillations involving
electron neutrinos are causing the  atmospheric neutrino anomaly.   This result is quantitatively
illustrated in Figure~\ref{fig:excl_summary}~\cite{apollonio}.

\section{Physics with KamLAND}

\begin{figure}[htbp]
\begin{center}
\fbox{\epsfig{file=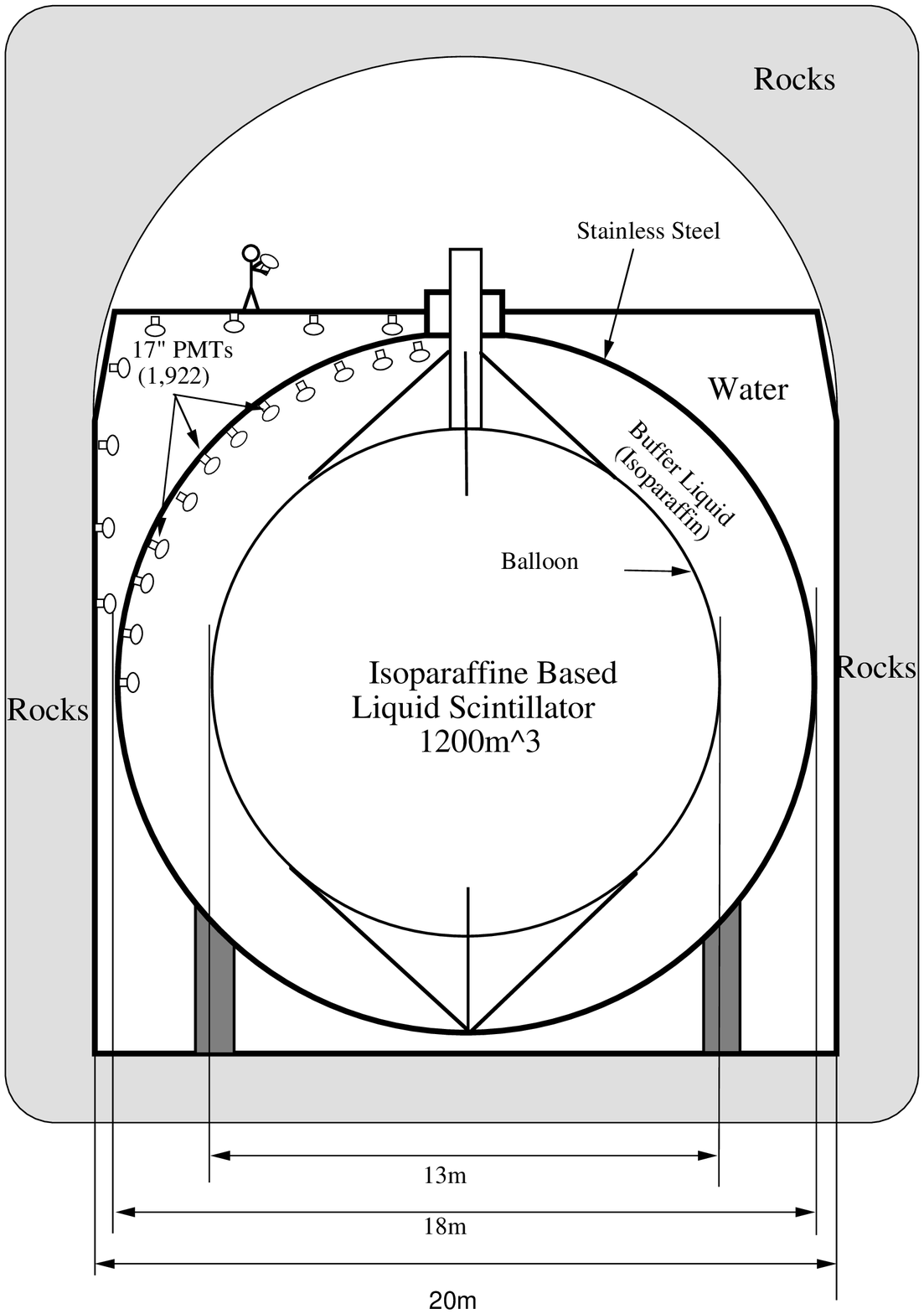,height=9cm,clip=}}
\caption{Schematic view of the KamLAND detector.}
\label{fig:kamland}
\end{center}
\end{figure}

The KamLAND experiment will use the Kamiokande  infrastructure, under 1000~m rock
overburden, to perform an ultra-long baseline experiment with enough sensitivity 
to test the large mixing angle MSW solution of the solar neutrino puzzle.
KamLAND, will consist of 1000~tons of liquid scintillator surrounded by 2.5~m thick
mineral oil shielding layer.   Both liquids
are contained in a 18~m diameter stainless-steel sphere that also supports, on the
inside surface, about 2000 17-inch photomultipliers giving a 30\% photocathode
coverage.    Such photomultipliers are modified from the 20-inch SuperKamiokande
tubes and provide 3~ns FWHM transit-time-spread, allowing 1~MeV energy 
depositions to be localized with 10~cm accuracy.    The detector light yield will
be better than 100~p.e./MeV.
A veto detector will be provided by flooding with water the volume outside the 
sphere and reading the \v{C}erenkov light with old Kamiokande photomultipliers.
A schematic view of the  detector is shown in Figure~\ref{fig:kamland}.  
In Table~\ref{tab:kl_rates} we list the power, distance and neutrino rates for the
five nuclear plants giving the largest $\bar\nu_{\rm e}$ flux contributions together 
with the total from all Japanese reactors ($\sim$2~events day$^{-1}$).
\begin{table}[htbp]
\begin{center}
\begin{tabular}{|l|r|r|r|r|}
\hline
  Reactor Site       & Number of &   Thermal Power    &     Distance     &    Rate       \\
                     & Reactors  &      (GW)          &       (km)       & (Events/year) \\
\hline
\hline
  Kashiwazaki        &    7      &     24.6           &       160        &    348        \\
  Ohi                &    4      &     13.7           &       180        &    154        \\
  Takahama           &    4      &     10.2           &       191        &    102        \\
  Hamaoka            &    4      &     10.6           &       214        &     84        \\
  Tsuruga            &    2      &      4.5           &       139        &     84        \\
\hline
  Total              &   51      &    130             &                  &   1075        \\
\hline
\end{tabular}
\renewcommand{\arraystretch}{1.0}
\vspace{10pt}
\caption{\footnotesize Expected contribution of different reactors to the 
neutrino rates to be detected in KamLAND in the case of no oscillations.  Several
other reactors each giving a small contribution do not have individual
entries in the table but are included in the total.   While the rates in the
table are referred to the nominal power of the reactors the typical running
duty cycle is 80\%.}
\label{tab:kl_rates}
\end{center}
\end{table}

\begin{figure}[htbp]
\begin{center}
\fbox{\epsfig{file=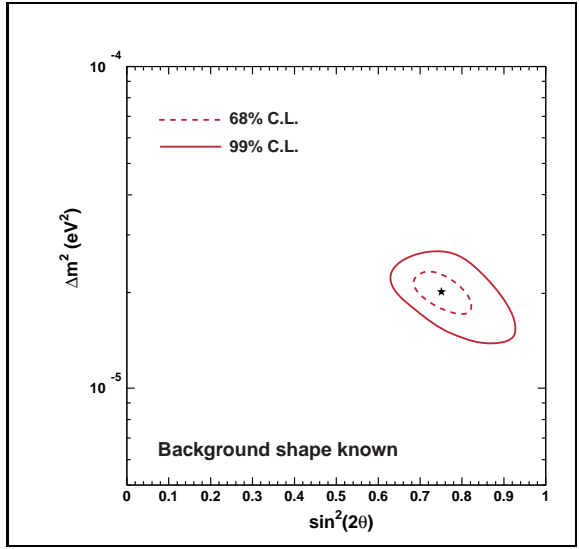,height=7cm,clip=}}
\caption{Error contours expected at 68\% and 99\% CL.   The plot assumes
3 years of KamLAND reactor data and oscillations at $\sin^2 2\theta = 0.75$, 
$\Delta m^2 = 2\times 10^{-5}$.  In the analysis the shape of the background 
in energy was assumed known but the integral was left as free parameter
in a fit to power excursions of the type shown in Figure~\ref{fig:flux_mod}.}
\label{fig:kl_evidence}
\end{center}
\end{figure}
Extensive detector simulations predict that the main backgrounds will result
from random coincidence of hits from natural radioactivity (0.05 events/day)
and neutrons produced by muon spallation in rock (0.05 events/day).   Hence
the signal/noise ratio for reactor anti-neutrinos is expected to be about 20/1.
While the predicted exclusion contour for the case of 3-years of data taking and no 
evidence of oscillations is shown in Figure~\ref{fig:excl_summary}, Figure~\ref{fig:kl_evidence}
shows the precision to which the two oscillation parameters  would be measured
in 3 years if oscillations would indeed occur according to the large mixing angle MSW 
solution of the solar neutrino puzzle.
In addition to the neutrino oscillation physics described above KamLAND will
also perform a number of new measurements in the fields of terrestrial 
neutrinos, supernovae physics and solar neutrinos.

Construction for KamLAND has started in 1998 and data taking is 
scheduled to begin during 2001.

In conclusion it appears that the study of reactor neutrinos is a very interesting field
indeed, offering the opportunity of exciting measurements and discoveries.
The next 5 to 10 years should be rich of results !

\section{Acknowledgments}

I would like to express my gratitude to my collaborators in the Palo Verde 
and KamLAND experiments for lots of interesting discussions on neutrino physics.
I am also indebted with C. Bemporad who
provided me with the material about the Chooz experiment.


\begin{thebibliography}{99}

\bibitem{ysuzuki} Y. Suzuki, these proceedings.

\bibitem{Louis} H. White, these proceedings.

\bibitem{chooz} H. de Kerret {\it et al.} ``Chooz Proposal'', 1993, Unpublished.

\bibitem{paloverde} F. Boehm {\it et al.} ``Proposal for the San Onofre 
                    Neutrino  Oscillation Experiment'', Caltech 1994, Unpublished; 
                    updated by F. B\"ohm {\it et al.} ``Addendum to the San Onofre 
                    Proposal'' Calt-63-721 (1996), Unpublished.

\bibitem{kamland} P. Alivisatos {\it et al.} ``KamLAND a Liquid scintillator Anti-Neutrino
                  Detector at the Kamioka site'', Stanford-HEP-98-03, Tohoku-RCNS-98-15, 
                  July 1998 (Unpublished).

\bibitem{schre} K. Schreckenbach {\it et al.}, 
                Phys. Lett. B99 (1981) 251; B160 (1985) 325; \\
                F. von Feilitzsch {\it et al.}, 
                Phys. Lett. B118 (1982) 162; \\
                A. A. Hahn {\it et al.}, 
                Phys. Lett. B218 (1989) 365.

\bibitem{oldexp} B. Achkar {\it et al.}, Nucl. Phys. B434 (1995) 503; \\
                 Y. Declais {\it et al.}, Phys. Lett. B338 (1995) 383; \\
                 G. Zacek {\it et al.}, Phys. Rev. D34 (1986) 2621.              

\bibitem{apollonio} M. Apollonio {\it et al.}, Phys. Lett. B420 (1998) 397.
          
\end{thebibliography}
\end{document}